\documentclass[12pt,draftclsnofoot, onecolumn]{IEEEtran}

\usepackage{amsmath}
\usepackage{mathrsfs}
\usepackage{subfigure}
\usepackage{multirow}
\usepackage[table,xcdraw]{xcolor}
\usepackage{booktabs}
\usepackage{amsfonts,amssymb} 
\usepackage{url}
\usepackage{flushend}
\usepackage[table,xcdraw]{xcolor}
\usepackage{algorithmic,algorithm}
\usepackage[noadjust]{cite}
\usepackage{cases}
\usepackage[table,xcdraw]{xcolor}
\usepackage{overpic}
\usepackage{textcomp}
\usepackage{graphicx,amssymb,lineno}
\usepackage{amsmath,amsfonts,amssymb}

\usepackage{subeqnarray}
\usepackage{cases}
\newtheorem{theorem}{Theorem}
\newtheorem{lemma}{Lemma}

\newtheorem{remark}{Remark}

\newcommand\wb{\ensuremath{{\boldsymbol w}}}

\newcommand\xb{\ensuremath{{\boldsymbol x}}}

\ifCLASSINFOpdf
\else
\fi

\begin{document}

\title{Low-complexity and Low-overhead Receiver for OTFS via Large-scale Antenna Array}
\author{Yaru Shan and Fanggang Wang

}


\maketitle
\begin{abstract}
Orthogonal time frequency space (OTFS) is a modulation technique that is dedicated to the high-speed mobility scenario. However, its transmission involves a two-dimensional convolution of the symbols of interest and the multipath fading channel, and it complicates the equalization. In addition to the high-complexity issue, the existing pilot pattern to estimate the unknown channel accurately requires large overhead to avoid the pilot being contaminated, which is spectrally inefficient. In this paper, we propose a receiver approach by the marriage of the OTFS and a large-scale antenna array, which allows low-complexity detection and low-overhead pilot pattern design. The receiver is briefly summarized as follows. First, the received signal from each path of the multipath fading channel is identified by a high-resolution receive beamformer facilitated by a large-scale antenna array. Then the identified signal from each angle in the delay-Doppler domain reduces to a flat-faded signal, which can be simply equalized using the channel information estimated by our pilot pattern. Moreover, the derivation shows that the received signal from an angle of arrival turns out to be a flat-faded signal with rotations in both the delay and the Doppler coordinates. We further provide the estimator of the channel fading and the rotations of delay and Doppler. With these estimates, the symbols of interest can be recovered, and then, the signals from all angles of arrival are combined as different diversity versions. In addition, our pilot pattern with only around $25$\% overhead of the existing pilot pattern ensures the same protection of pilot pollution. The significance of the proposed receiver is its practicality, and it achieves better error performance with lower receiver complexity and lower overhead compared to the existing approaches, at the cost of a linear beamforming antenna array. The price is quite affordable, since the linear antenna array has moderate computational complexity, and it is deployed widely in current and future wireless communication systems.
Eventually, the efficiency, the reliability, and the low complicacy of the proposed receiver approach are further validated by the numerical results.
\end{abstract}
\section{Introduction}
Future mobile communications are expected to support numerous large-throughput applications in high mobility scenarios such as high-speed railway, highway, and unmanned aerial vehicles \cite{background,highspeed3,highspeed1,highspeed2}. However, the multipath propagation and the Doppler effect cause the channel dispersion in both the time domain and the frequency domain, which degrade the error performance, especially in a high mobility environment. Currently, orthogonal frequency division multiplexing (OFDM) is sufficiently studied in the time-dispersive channel. It eliminates the inter-symbol interference (ISI) while suffering from severe inter-carrier interference (ICI) which can be caused by the Doppler effect \cite{performanceofOFDM}. An alternative modulation scheme is exigent to resist the rapid channel time variation.

Recently, a new two-dimensional multicarrier modulation scheme called orthogonal time frequency space (OTFS) has been proposed in \cite{OTFSAppear1,OTFSAppear2, OTFSandHS} to combat fast time-varying channels with high Doppler effect. The OTFS technique has an intrinsic property of eliminating the need to adapt to the fast varying channel \cite{OTFSAppear2}. The OTFS utilizes the delay-Doppler domain, instead of the frequency-time domain, to multiplex the information-bearing symbols. The fast-varying channel in the frequency-time domain is converted into the one in the delay-Doppler domain such that all symbols in a transmission frame experience an almost static channel. However, the equivalent transmission in the delay-Doppler domain involves a sophisticated two-dimensional periodic convolution, and it makes the practicality of the OTFS system quite challenging.

Low-complexity equalization for the OTFS has attracted much attention \cite{DoneCEandEQ,EQ1,EQ2,EQ4,EQ5,EQ6,EQ7}. To avoid the two-dimensional convolution in the delay-Doppler domain, some work adopts the domain transformation to simplify the equalization. In \cite{EQ1}, the equalization conducts in the conventional frequency domain first and then resorts to the delay-Doppler domain to mitigate the residual interference. This method sacrifices the reliability of the OTFS in the high-speed scenario to reduce the complexity of the detection.
In \cite{EQ2}, minimum mean square error-parallel interference cancellation was also used in the frequency-time domain to guarantee the reliability, however, the performance was still limited due to the Doppler effect.
Moreover, there is some work conducting the equalization in the delay-Doppler domain.
In \cite{DoneCEandEQ}, Gibbs sampling was employed to get an approximated maximum likelihood solution. However, the scheme did not take advantage of the sparsity of the channel in the delay-Doppler domain \cite{lj}.
Based on a sparse factor graph and the Gaussian approximation of the interference, a low-complexity message passing (MP) detection of the uncoded OTFS was proposed in \cite{EQ4} and \cite{EQ5}. The complexity of this method
relies on the sparsity of the channel and varies in different channel models. In the low signal to noise ratio (SNR) region, the complexity of the MP increases sharply due to  the emergence of the loopy graph resulting in poor performance.
In \cite{EQ6}, an iterative detection algorithm was proposed to reduce the complexity by adopting the first order Neumann series
to approximate the involved matrix inversion.
In addition, the low-complexity detection algorithm in \cite{EQ7} relied on the assumption that each path has almost the same Doppler effect limiting the application.

Channel information is required to detect the symbols in the delay-Doppler domain in the OTFS. The choice of domain affects the complexity of channel estimation. The authors in \cite{DoneCEandEQ} and \cite{TITCE1} proposed a channel estimation method in the frequency-time domain, however, with high complexity. In \cite{CE3}, a whole frame was used as the pilot in the delay-Doppler domain. However, the channel estimate in the first frame was used for the following several frames, which deteriorates the error performance in a fast-varying channel environment. In \cite{CE4}, the pilot, data symbols, and the guard band were carefully allocated to estimate the channel state in the delay-Doppler domain. However, a large number of guard symbols are required to prevent the pilot from being polluted by the ambient data symbols, which reduces the spectral efficiency.

In this paper, we design a low-overhead and low-complexity receiver scheme for the OTFS system, including the pilot pattern design, the channel estimation, and the symbol detection. In particular, a large-scale antenna array is deployed at the receiver to decouple the received signals from different angles of arrival (AoA) into multiple parallel signal branches. By a spatial matched filter realized by a receive beamformer, the received signal from each path of the multipath channel is identified. The two-dimensional convolution in the delay-Doppler domain is decoupled. The subsequent channel estimation and the equalization conduct in each identified path. For channel estimation, we claim that only around $25$\% overhead of the existing pilot pattern is sufficient to obtain the same level of protection hence improving the spectral efficiency. For each identified path, the channel fading and the shifts in the delay and the Doppler coordinates can be accurately estimated by our pilot pattern.
In the equalization, after compensating for the delay and the Doppler shifts, the received signal can be equalized as a flat-faded signal reducing the complexity of the symbol detection. Lastly, the received signals with different AoAs are maximal-ratio combined to determine the information-bearing symbols. The main contributions of this paper are summarized as follows:

\begin{itemize}
\item Proposed an OTFS receiver scheme. By deploying a large-scale antenna array at the receiver, the received signals from different AoAs are decoupled into multiple parallel branches through the receive beamforming. The input-output relation in the delay-Doppler domain of each identified path is derived, which can be regarded as a flat-faded signal.

\item Designed a pilot pattern and proposed a low-overhead channel estimation method. Compared with the pilot pattern in \cite{CE4}, our pilot overhead almost reduces to around $25$\% of that in \cite{CE4}. Furthermore, the accuracy of the channel estimation is improved compared to the existing ones.

\item The proposed detection is low-complexity and non-iterative in contrast to the MP symbol detection, especially in the low-SNR region. The main reason is that we treat the signal in each identified path individually and then collect all the signals by the maximal-ratio combining (MRC).
\end{itemize}

The remainder of the paper is organized as follows: the preliminaries of the OTFS modulation are introduced in Section II. Section III depicts our low-overhead and low-complexity receiver. Then, Section IV demonstrates numerical results, and it is followed by the conclusion in Section V.

\emph{Notation:}
Throughout this paper, variables and vectors are written as italic letters $x$ and bold italic letters $\xb$, respectively. $f(\cdot,\cdot)$ and $f[\cdot,\cdot]$ represent functions with continuous and integer arguments, respectively.
The operators $(\cdot)^{\ast}$, $[\cdot]_{M}$, and $\lfloor \cdot \rfloor$ denote the complex conjugate, the modulo $M$, and the floor of the argument. The operation $\circledast$ represents the two-dimensional convolution. Let $j=\sqrt{-1}$, and $\delta(\cdot)$ is the Dirac delta function. Define $\mathcal{I}_{N}=\{0, 1, \dots, N-1\}$ as shorthand hereafter to represent an index set. $|\mathcal{X}|$ is the cardinality of the set $\mathcal{X}$.

\section{Preliminaries for OTFS}
In this section, we first introduce the OTFS delay-Doppler domain representation and compare the signal grid therein with the one in the frequency-time domain. Then, the modulation and the demodulation for the OTFS are previewed. Finally, the input-output relation of the OTFS system in the delay-Doppler domain is illustrated.
\subsection{Delay-Doppler Signal Representation}
The delay-Doppler signal representation is one fundamental signal representation that traces back to the work in \cite{delay-Dopplersignal}. A signal in the delay-Doppler domain is a function of $\phi(\tau,\nu)$ satisfying the quasi-periodicity condition \cite{OTFSAppear2} as
\begin{align}
\phi(\tau+\rho\tau_{\rm{r}}, \nu+\varrho\nu_{\rm{r}})=e^{j2\pi(\rho\nu\tau_{\rm{r}}-\varrho\tau\nu_{\rm{r}})}\phi(\tau, \nu)
\label{eq:1}
\end{align}
where $\tau$ and $\nu$ are the delay and the Doppler variables, respectively; $\tau_{\rm{r}}$ and $\nu_{\rm{r}}$ are the periods of the delay and the Doppler, and $\rho$ and $\varrho$ are the numbers of their traversal, respectively. In addition, $\tau_{\rm{r}}$ and $\nu_{\rm{r}}$ determine the delay-Doppler domain representation since a continuous delay-Doppler representation family is associated with the pair parameter $(\tau_{\rm{r}},\nu_{\rm{r}})$ in the hyperbola $\tau_{\rm{r}}=\frac{1}{\nu_{\rm{r}}}$.
The information-bearing symbols are multiplexed into a constant channel in the delay-Doppler domain. The channel-symbol coupling, i.e., two-dimensional periodic convolution, has three remarkable properties: invariance, separability, and orthogonality which can be exploited in the channel estimation and the equalization for the OTFS \cite{OTFSAppear2}.

\subsection{Basic Concepts in Signal Grids}
The delay-Doppler grid and the frequency-time grid are shown in Figure $1$ and the comparison between the two grids is described as follows.
\subsubsection{Delay-Doppler grid}
\begin{figure*}[tbp]
\centering
\centerline{\includegraphics[width=1.0\textwidth]{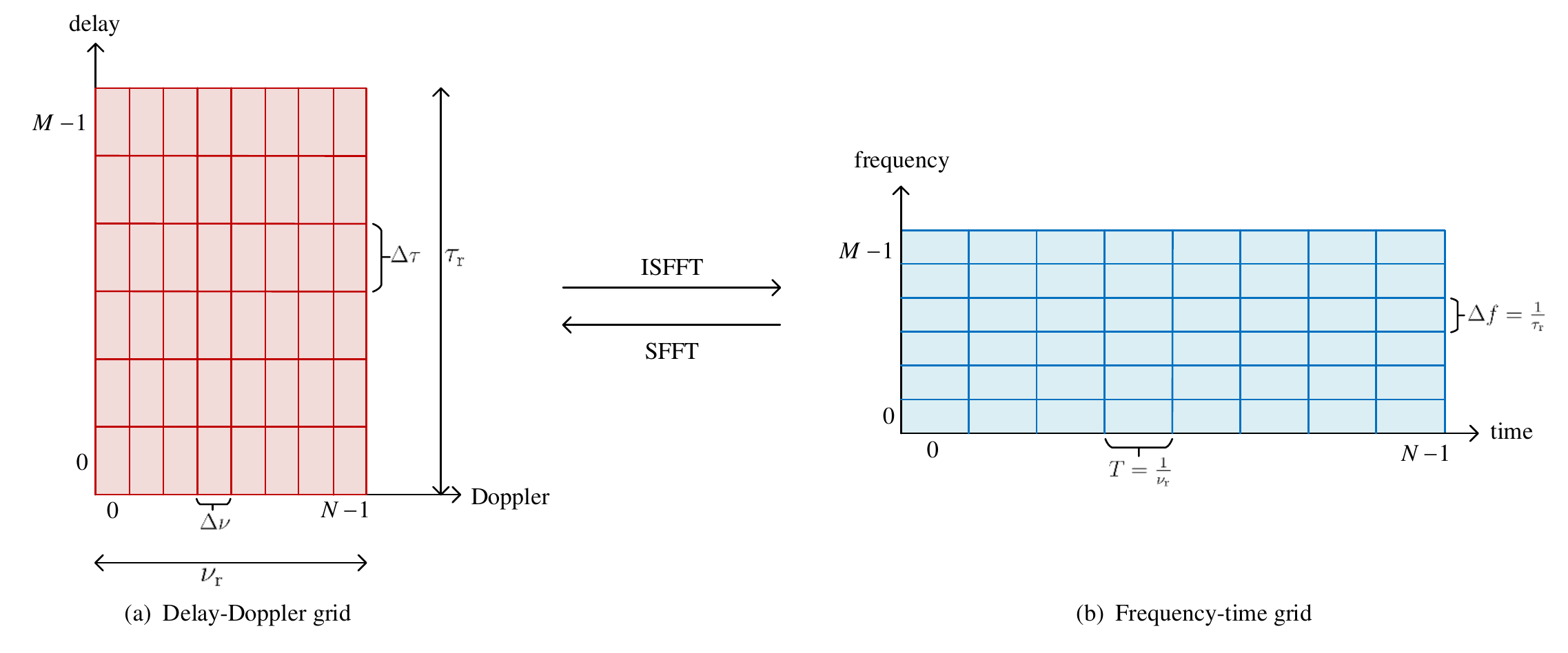}}
\caption{The delay-Doppler grid (left) and the frequency-time grid (right) when $M=6$ and $N=8$ in this example.}
\label{fig}
\end{figure*}
The delay-Doppler grid is defined in the rectangular unit if the delay and the Doppler periods are determined.
The two-dimensional grid in the delay-Doppler domain consists of $M$ points along the delay coordinate with the spacing $\Delta\tau$, and it has $N$ points along the Doppler coordinate with the spacing $\Delta\nu$, resulting in a set of $MN$ grid points inside the fundamental rectangular domain, which is expressed as
\begin{align}
\mathcal{F}=\left\{\left(l\Delta\tau, k\Delta\nu\right),\quad l\in \mathcal{I}_{M},\ k\in\mathcal{I}_{N} \right\}
\end{align}
where $\Delta\tau=\frac{\tau_{\rm{r}}}{M}$ and $\Delta\nu=\frac{\nu_{\rm{r}}}{N}$.
A wireless doubly-dispersive channel in the delay-Doppler domain is assumed to be bounded by the finite support $[0,\tau_{\text{max}}]$  along the delay coordinate and be bounded by $[-\nu_{\text{max}}, \nu_{\text{max}}]$ along the Doppler coordinate. $\tau_{\text{max}}$ and $\nu_{\text{max}}$ are the maximum delay spread and the maximum Doppler shift of the channel and in general $\tau_{\text{max}}\nu_{\text{max}}\ll1$. $\tau_{\text{r}}$ and $\nu_{\text{r}}$ are chosen to satisfy $\tau_{\text{r}}\gg\tau_{\text{max}}$ and $\nu_{\text{r}}\gg\nu_{\text{max}}$
which is enabled due to the fact that $\tau_{\text{r}}\nu_{\text{r}}=1$.

\subsubsection{Frequency-time grid}
\begin{figure*}[tbp]
\centering
\centerline{\includegraphics[width=1.0\textwidth]{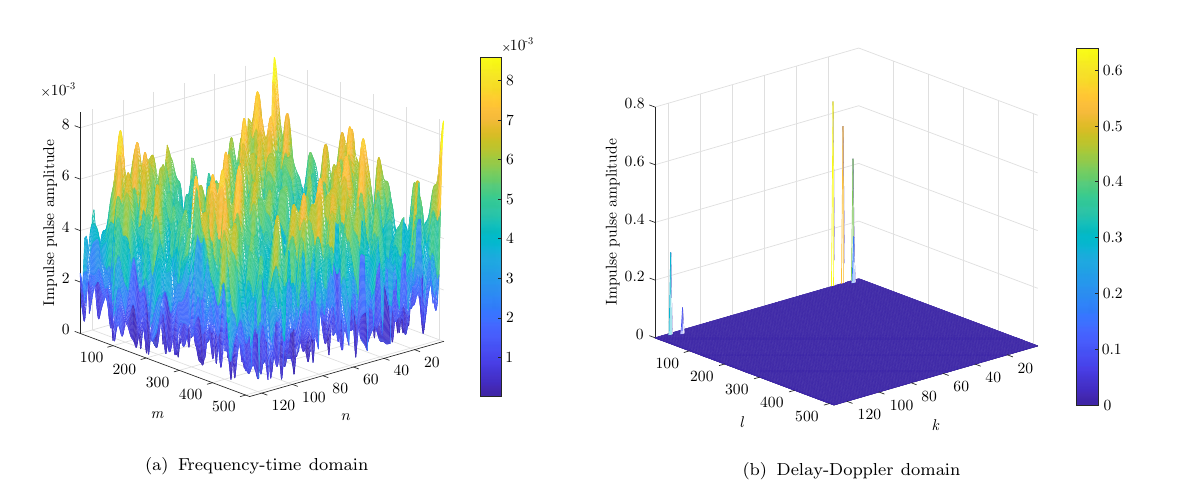}}
\caption{The amplitude of the impulse response of the Extend Vehicular A model \cite{channelEVA} in (a) the frequency-time grid, and (b) the delay-Doppler grid. The channel response in the delay-Doppler domain exhibits a sparse characteristic compared to the channel in the frequency-time domain.}
\label{fig}
\end{figure*}
The corresponding two-dimensional frequency-time domain grid consists of $M$ points along the frequency coordinate with the spacing $\Delta f$, and $N$ points along the time coordinate with the spacing $T$ which is expressed as
\begin{align}
\mathcal{G}=\left\{(m\Delta f,nT),\hspace{1em} m\in \mathcal{I}_{M},\ n\in\mathcal{I}_{N}\right\}
\end{align}
where the subcarrier spacing $\Delta f=\frac{1}{\tau_{\rm{r}}}$ and the symbol duration $T=\frac{1}{\nu_{\rm{r}}}$.
Comparing the resolution of the delay-Doppler and the frequency-time domain grids, the relation between the resolutions along the delay and the time can be expressed as
\begin{align}
T&=\tau_{\text{r}}\\
&=M\Delta\tau.
\end{align}
The relation between the resolutions along the frequency and the Doppler can be written as
\begin{align}
\Delta f&=\nu_{\text{r}}\\
&=N\Delta\nu.
\end{align}
In other words, the delay-Doppler is a fine-granularity domain compared with the frequency-time domain. The channel response in the delay-Doppler domain exhibits the sparse and slow-varying characteristics. For an explicit illustration, we mesh the amplitude of the impulse response of the Extend Vehicular A model \cite{channelEVA} in the speed of $500$ km/hr in Figure $2$. We can see that the channel responses in the frequency-time domain vary rapidly than the ones in the delay-Doppler domain. Note that the channel in the delay-Doppler domain is the symplectic finite Fourier transform (SFFT) of that in the frequency-time domain, and it sparsely locates in the delay-Doppler points. The sparsity property is crucial which can be exploited in the channel estimation and the equalization.

\begin{figure*}[tbp]
\centering
\centerline{\includegraphics[width=1\textwidth]{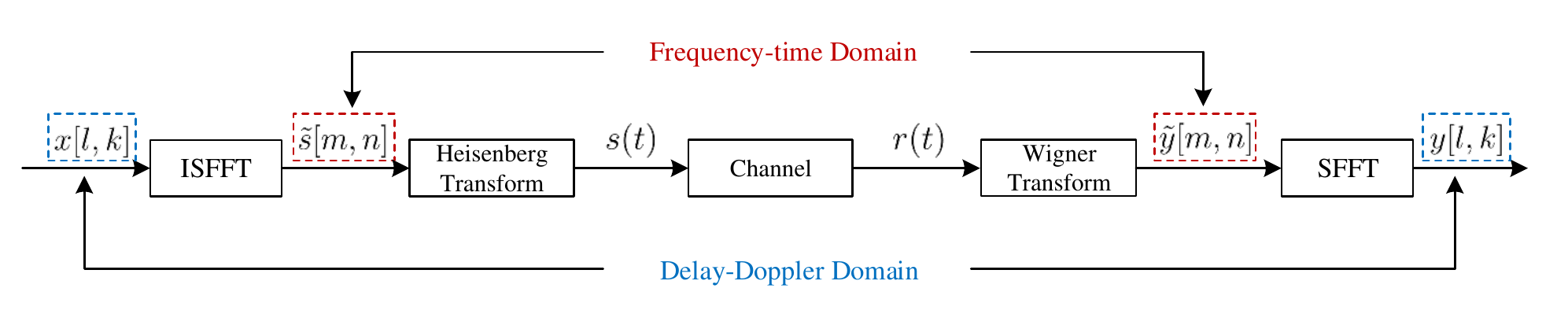}}
\caption{The modulation and the demodulation diagram of OTFS.}
\label{fig}
\end{figure*}
\subsection{OTFS Modulation and Demodulation}
The modulation and demodulation diagram of the OTFS is shown in Figure $3$. First, the $MN$ information-bearing symbols represented by $x[l,k]$, $l\in\mathcal{I}_{M}$, $k\in\mathcal{I}_{N}$, are mounted into the delay-Doppler domain. By utilizing the inverse SFFT (ISFFT), the $MN$ symbols are transformed into the frequency-time domain as
\begin{align}
\tilde{s}[m,n]=\frac{1}{\sqrt{MN}}\sum_{l=0}^{M-1}\sum_{k=0}^{N-1}x[l,k]e^{j2\pi\left(\frac{nk}{N}-\frac{ml}{M}\right)},\qquad m\in\mathcal{I}_{M},\ n\in\mathcal{I}_{N}.
\label{eq:modu1}
\end{align}
The specific coupling $\frac{nk}{N}-\frac{ml}{M}$ of the exponent explains the symplectic term. Then, the Heisenberg transform is adopted to transform the signal into the time domain. By appending a cyclic prefix (CP) of a length $t_{\text{cp}}$, the time-domain signal to be sent can be expressed as
\begin{align}
s(t)=\sum_{m=0}^{M-1}\sum_{n=0}^{N-1}\tilde{s}[m,n]{g}_{\text{t}}(t-nT)e^{j2\pi m\Delta f(t-nT)},\qquad t\in(-t_{\text{cp}}, NT)
\label{eq:modu2}
\end{align}
where $g_{\text{t}}(\cdot)$ denotes the transmit pulse shaping function.

The received signal is a superposition of the multiple reflected replicas of the transmitted signal, where each replica lags in the path delay, frequency-deviates by the Doppler shift, and is weighted by the time-independent complex-valued delay-Doppler impulse response $h(\tau,\nu)$. After removing the CP, the received signal can be expressed as
\begin{align}
r(t)=\int_{-\nu_{\text{max}}}^{\nu_{\text{max}}}\int_{0}^{\tau_{\text{max}}} h(\tau,\nu)s(t-\tau)e^{j2\pi \nu(t-\tau)}{\rm d}\tau {\rm d}\nu, \qquad t\in(0, NT).
\label{eq:modu3}
\end{align}
At the receiver, the output of the receive filter is obtained as
\begin{align}
\tilde{y}[m,n]=\int_{t} g^{\ast}_{\text{r}}(t-nT)r(t)e^{-j2\pi m\Delta f(t-nT)}{\rm d}t,\qquad m\in\mathcal{I}_{M},\ n\in\mathcal{I}_{N}
\label{eq:modu5}
\end{align}
where $g_{\text{r}}(\cdot)$ denotes the receive pulse shaping function.
The operation is referred to as Wigner transform \cite{EQ4} which transforms the received signal from the time domain to the frequency-time domain.
The received signal is then transformed into the delay-Doppler domain by the SFFT, which is expressed as
\begin{align}
y[l,k]=\frac{1}{\sqrt{MN}}\sum_{m=0}^{M-1}\sum_{n=0}^{N-1}\tilde{y}[m,n]e^{-j2 \pi\left(\frac{nk}{N}-\frac{ml}{M}\right)},\qquad l\in \mathcal{I}_{M}, \ k\in\mathcal{I}_{N}.
\label{eq:modu6}
\end{align}
\subsection{Input-output Relation of OTFS Transmission}
The delay-Doppler domain channel model and the input-output relation of the OTFS-based communication system are described as follows.
The channel responses are modeled as $P$ different delay taps. For each delay tap, there could be multiple different Doppler shifts, since there might exist multiple scatters located on the elapse focused by both the transmitter and the receiver. Thus, the delay-Doppler channel is modeled as
\begin{align}
h(\tau,\nu)=\sum_{p=0}^{P-1}\sum_{q=0}^{Q_{p}-1}\beta_{p,q}\delta(\tau-\tau_{p})\delta(\nu-\nu_{p,q})
\label{eq:ch1}
\end{align}
where $\tau_{p}$ is the delay for the $p$th tap; $Q_{p}$ is the number of the paths having the identical delay $\tau_p$; $\nu_{p,q}$ is the Doppler shift of the $q$th path in the $p$th tap; $\beta_{p,q}$ is the normalized complex-valued response of the $q$th path in the $p$th tap.
The continuous delay $\tau_{p}$ and the Doppler shift $\nu_{p,q}$ can be discretized as
\begin{align}
l_{p}&=M\Delta f\tau_{p},\qquad p\in\mathcal{I}_{p} \\
k_{p,q}&=NT\nu_{p,q},\qquad p\in\mathcal{I}_{p},\ q\in\mathcal{I}_{Q_{p}}
\label{eq:ch2}
\end{align}
where $l_{p}$ and $k_{p,q}$ are the delay and Doppler indices, respectively.\footnote{As the delay resolution is high such that we ignore the approximation. Similarly, we ignore the approximation of the fractional Doppler shift since the high resolution is provided by the large-scale antenna array which is further demonstrated in Section III.}

The input-output relation of the delay-Doppler domain is first derived in \cite{EQ5}. It can be simplified when the ISI and ICI are removed assuming that the transmit pulse $g_{\text{t}}(t)$ and the receive pulse $g_{\text{r}}(t)$ satisfy the bi-orthogonality condition \cite{EQ5}. The cross-ambiguity function of the transmit and the receive pulse is defined as
\begin{align}
A_{g_{\text{r}},g_{\text{t}}}(f,t)=\int_{t'} g_{\text{r}}^{\ast}(t'-t)g_{\text{t}}(t')e^{-j2\pi f(t'-t)}{\rm d}t'.
\label{eq:cross}
\end{align}
Based on the cross-ambiguity function, the bi-orthogonality condition of the $g_{\text{t}}(t)$ and $g_{\text{r}}(t)$ is expressed as
\begin{align}
A_{g_{\text{r}},g_{\text{t}}}(f,t)|_{f=m\Delta f+(-\nu_{\text{max}}, \nu_{\text{max}}),\ t=nT+(-\tau_{\text{max}}, \tau_{\text{max}})}
=\delta[m]\delta[n]u_{\nu_{\text{max}}}(f)u_{\tau_{\text{max}}}(t)
\label{eq:ch3}
\end{align}
where $u_{c}(x)=1$ for $x\in(-c,c)$ and $u_{c}(x)=0$ otherwise.\footnote{Currently, the bi-orthogonal pulses can not be realized in practice. However, the pulses can be approximated by the pulses whose support is highly concentrated in both the frequency and the time dimensions \cite{Sparsity}.}
We simply assume that the inter-Doppler interference (IDI) is trivial which can be ignored when the delay-Doppler grid is fine-granularity. Without the ICI, the ISI and the IDI, the input-output relation in the delay-Doppler domain can be expressed as
\begin{align}
y[l,k]&=h[l,k]\circledast x[l,k]\\
&=\sum_{p=0}^{P-1}\sum_{q=0}^{Q_{p}-1}\beta_{p,q}e^{-j2\pi\tau_{p}\nu_{p,q}}x\left[[l-l_{p}]_{M},[k-k_{p,q}]_{N}\right], \qquad l\in \mathcal{I}_{M},\ k\in\mathcal{I}_{N}
\label{eq:io2dim}
\end{align}
where $h[l,k]$ is obtained by sampling $h(\tau,\nu)$ at $l= M\Delta f\tau$ and $k=NT\nu$. We can see that each symbol in the delay-Doppler domain is spread by all delay taps and all Doppler shifts due to the two-dimensional convolution, which is quite involved in the channel estimation and the symbol detection. In the following, we introduce the proposed low-complexity receiver approach.

\section{Proposed Receiver Approach}
In this section, we first introduce the system model. Then, the receiver design is elaborated including the receive beamforming and the low-complexity detection algorithm. Finally, the low-overhead pilot pattern design and the channel estimation algorithm are introduced.
\subsection{System Model}
Consider a high-mobility downlink transmission where the base station transmits a signal to a mobile user. A large-scale uniform linear antenna array is configured at the receiver over its heading direction.

A set of $MN$ pilot, guard, and information-bearing symbols $x[l,k]$ are multiplexed in the delay-Doppler grid $\mathcal{F}$. Without loss of generality, we assume the transmit power is normalized. By the ISFFT in (\ref{eq:modu1}) and the Heisenberg transformation in (\ref{eq:modu2}), the signal is transformed into the time domain signal $s(t)$.
We consider a multipath channel model from the base station to the $i$th receive antenna, which is expressed as
\begin{align}
h_{i}(t,\tau)=\sum_{p=0}^{P-1}\sum_{q=0}^{Q_{p}-1}\beta_{p,q}e^{j(2\pi f_{\text{d}}t+\phi_{i})\cos\theta_{p,q}}\delta(\tau-\tau_{p}), \qquad i \in\mathcal{I}_{E}
\label{eq:h1}
\end{align}
where $E$ is the number of the receive antennas; the maximum Doppler shift $f_{\text{d}}$ is defined as $f_{\text{d}}=\frac{v}{\lambda}$, $v$ is the velocity of the mobile user, and $\lambda$ is the carrier wavelength; $\theta_{p,q}$ is the AoA of the $q$th path in the $p$th tap. The phase of a receive antenna is expressed as
\begin{align}
\phi_{i}=\frac{1}{\lambda}2\pi i\eta, \qquad i\in\mathcal{I}_{E}
\label{eq:h3}
\end{align}
where $\eta$ is the antenna distance of the uniform array.
The received signal is represented by
\begin{align}
\tilde{r}_{i}(t)=\sum_{p=0}^{P-1}\sum_{q=0}^{Q_{p}-1}\beta_{p,q}e^{j(2\pi f_{\text{d}}t+\phi_{i})\cos\theta_{p,q}}s(t-\tau_{p}) +\tilde{z}_{i}(t), \qquad i\in\mathcal{I}_{E}
\label{eq:rr2}
\end{align}
where $\tilde{z}_{i}(t)$ is a circularly symmetric complex Gaussian (CSCG) noise at the $i$th receive antenna and it follows $\mathcal{CN}(0, \sigma^{2})$ at a time instant.
\subsection{Receiver Design}
In this subsection, we first adopt a beamformer at the receiver. Then, our equalization is elaborated.
\begin{figure*}[tbp]
\centering
\centerline{\includegraphics[width=1.0\textwidth]{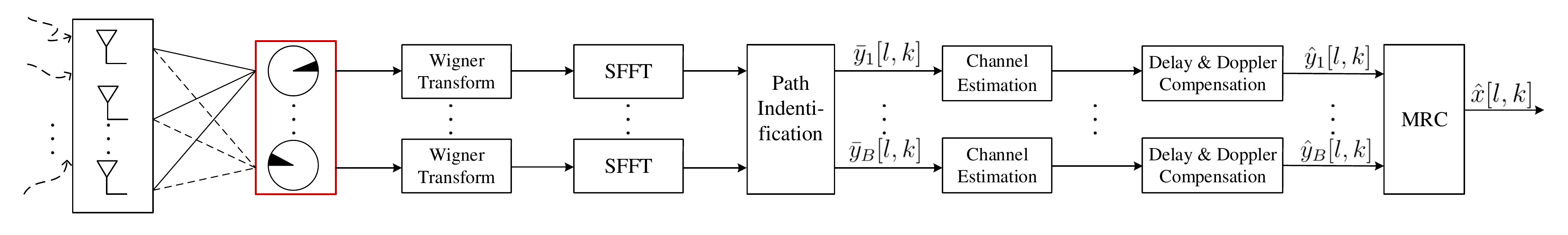}}
\caption{The diagram of the proposed receiver scheme.}
\label{fig}
\end{figure*}

\subsubsection{Receive beamforming}
The proposed receiver diagram is shown in Figure $4$. Since the Doppler shift can be identified by the AoA of each path, the receive beamforming is to extract the signal from the direction of interest while eliminating the signal from others. This is implemented by a spatial matched filter implemented by a receive beamformer. The steering vector of the uniform antenna array $\wb(\theta)$ is given by
\begin{align}
w_{i}(\theta)=e^{j\phi_{i}\cos\theta}, \qquad i\in\mathcal{I}_{E}.
\label{eq:steer}
\end{align}
To scan the possible AoAs of the multiple paths, we preset the matched angles over all directions. We assume that there are $B$ branches that receive the desired signal. In addition, the angles of interest are in the set $\varPhi_{B}=\{\varphi_{b}|b\in\mathcal{I}_{B}\}$. Furthermore, the $B$ paths are determined by thresholding the amplitude of the received signals of all directions. Define a one-to-one mapping function as $b=\varpi(p,q)$, $b\in\mathcal{I}_{B}$, where the index $(p,q)$ maps to the identified path $b$.
Thus, the received signal from the angle $\varphi_b$ is expressed as
\begin{align}
r_{b}(t)&=\frac{1}{E}\sum_{i=0}^{E-1}w^{\ast}_{i}(\varphi_{b})\tilde{r}_{i}(t) \\
&=\underbrace{\beta_{p,q}e^{j2\pi f_{\text{d}}t\cos\theta_{p,q}}s(t-\tau_{p})}_{\text{desired signal}}+\underbrace{z_{b}(t)}_{\text{noise}} \nonumber\\
&\quad+\underbrace{\frac{1}{E}\sum_{i=0}^{E-1}\sum_{\varpi(p',q')\neq b}\beta_{p',q'}e^{j2\pi f_{\text{d}}t\cos\theta_{p',q'}} e^{j\phi_{i}\left(\cos\theta_{p',q'}-\cos\varphi_{b}\right)}s(t-\tau_{p'})}_{\text{interference}},\quad\varpi(p,q)=b,\, b\in \mathcal{I}_{B}
\label{eq:bf4}
\end{align}
where
\begin{align}
z_{b}(t)=\frac{1}{E}\sum_{i=0}^{E-1}w^{\ast}_{i}(\varphi_{b})\tilde{z}_{i}(t), \qquad b\in \mathcal{I}_{B}.
\end{align}
The interference term in (\ref{eq:bf4}) is trivial for a large-scale antenna array,\footnote{Note that the signal from the angle of $2\pi-\varphi_{b}$ has the identical array gain as $\varphi_{b}$. However, this occasion occurs with probability $0$, and thus, it is ignored hereafter. More details about the asymptotic analysis are referred to as Appendix A.} and then the received signal can be approximated as
\begin{align}
r_{b}(t)\thickapprox\beta_{b}e^{j2\pi\nu_{b} t} s(t-\tau_{b})+z_{b}(t), \qquad b\in \mathcal{I}_{B}
\end{align}
where denote $\beta_{b}\triangleq \beta_{p,q}$, $\tau_b\triangleq \tau_{p}$, $b=\varpi(p,q)$ for convenience, and $\nu_{b}=f_{\text{d}}\cos\varphi_{b}$ is the Doppler shift of the $b$th identified path.
In addition, we can see that the channel response of the $b$th identified path reduces to a Dirac delta function at a single delay shift and a single Doppler shift which can be expressed in the delay-Doppler domain as
\begin{align}
h_{b}(\tau,\nu)=\beta_{b}\delta(\tau-\tau_{b})\delta(\nu-\nu_{b}), \qquad b\in\mathcal{I}_{B}.
\label{eq:dd1}
\end{align}
\subsubsection{Low-complexity detection}
The low-complexity detection conducts in the delay-Doppler domain. The received signal $r_{b}(t)$ is first transformed into the frequency-time domain $\tilde{y}_{b}[m,n]$ as (\ref{eq:modu5}), which is then transformed into the delay-Doppler domain by the SFFT as
\begin{align}
\bar{y}_{b}[l,k]&=\frac{1}{\sqrt{MN}}\sum_{m=0}^{M-1}\sum_{n=0}^{N-1}\tilde{y}_{b}[m,n]e^{-j2 \pi\left(\frac{nk}{N}-\frac{ml}{M}\right)},\qquad b \in\mathcal{I}_{B},\ l\in \mathcal{I}_{M},\ k\in\mathcal{I}_{N}.
\label{eq:b3}
\end{align}
The frequency-time domain input-output relation is given in Lemma $1$, and  then the relation in the delay-Doppler domain is provided in Theorem $1$.
\begin{lemma}
With the bi-orthogonality condition in (\ref{eq:ch3}), the input-output relation in the frequency-time domain can be expressed as
\begin{align}
\tilde{y}_{b}[m,n]=\beta_{b}e^{j2\pi nT\nu_{b}}e^{-j2\pi(\nu_{b}+m\Delta f)\tau_{b}}\tilde{s}[m,n], \qquad b \in\mathcal{I}_{B},\ m\in\mathcal{I}_{M}, \ n\in\mathcal{I}_{N}
\label{eq:io14}
\end{align}
\end{lemma}
\quad\quad $\emph{Proof:}$ See Appendix B.$\hfill\square$

\begin{theorem}
With the bi-orthogonality condition, the input-output relation in the delay-Doppler domain is expressed as
\begin{align}
\bar{y}_{b}[l,k]=\beta_{b}e^{-j2\pi \tau_{b}\nu_{b}}x[[l-l_{b}]_{M},[k-k_{b}]_{N}],\qquad b \in\mathcal{I}_{B}, \ l\in \mathcal{I}_{M}, \ k\in\mathcal{I}_{N}
\label{eq:io1}
\end{align}
where $l_{b}=M\Delta f\tau_{b}$ and $k_{b}=NT\nu_{b}$.
\end{theorem}
\quad\quad $\emph{Proof:}$ See Appendix C.$\hfill\square$
\begin{remark}
The input-output relation in the delay-Doppler domain indicates that the received symbols are rotated by the unified delay and Doppler shift. In addition, all received symbols in the identified path $\varphi_{b}$ are scaled by the gain of the identified path $|\beta_{b}|$, $b\in\mathcal{I}_{B}$, and phase-rotated by  $e^{j(\angle\beta_{b}-2\pi\tau_{b}\nu_{b})}$, $b\in\mathcal{I}_{B}$. This observation demonstrates that the received signal in each identified path can be regarded as a flat-faded signal. Since there are $B$ diversity branches, they can be maximal-ratio combined when all the branches are aligned with respect to the periodic delay and Doppler shifts.
\end{remark}

The equalization conducts as the following two steps: we compensate for the delay and Doppler shifts for each identified path and perform the MRC. First, the compensation is performed as
\begin{align}
\hat{y}_{b}[l,k]=\bar{y}_{b}\big[\big[l+l_{b}\big]_{M},\big[k+k_{b}\big]_{N}\big],\qquad b \in\mathcal{I}_{B}, \ l\in\mathcal{I}_{M}, \ k\in\mathcal{I}_{N}.
\label{eq:eq1}
\end{align}
Then the symbols of interest can be estimated by
\begin{align}
\hat{x}[l,k]=\frac{\sum_{b=0}^{B-1}\beta_{b}^{\ast}e^{j2\pi \tau_{b}\nu_{b}}\hat{y}_{b}[l,k]}{\sum_{b=0}^{B-1}|\beta_{b}|^{2}},\qquad l\in\mathcal{I}_{M}, \ k\in\mathcal{I}_{N}.
\label{eq:eq2}
\end{align}
\subsubsection{Comparison with MP detection}
The MP detection of the OTFS is adopted in \cite{EQ5}, \cite{CE4}, \cite{addmp1}, and \cite{addmp2}. The computational complexity of it is in the order of $O(n_{\mathsf {iter}}\sum_{p}Q_{p}MNC)$, where $n_{\mathsf {iter}}$ is the number of the iteration, and $C$ represents the size of the modulation alphabet. The complexity of the MP is affected by the sparsity of the channel \cite{EQ5}. In addition, the value of the $n_{\mathsf{iter}}$ increases sharply for the low SNR region with the emergence of the loopy graph.
Compared with MP, the complexity analysis of our proposed symbol detection is as follows. The complexities of (\ref{eq:eq1}) and (\ref{eq:eq2}) are in the order of  $O(MN)$ and $O(BMN)$, respectively. In addition, consider the further step of demodulation, the total complexity of our symbol detection is in the order of $O(BMNC)$. In general, $B$ is lower than $n_{\mathsf {iter}}\sum_{p}Q_{p}$ due to $B=\sum_{p}Q_{p}<n_{\mathsf {iter}}\sum_{p}Q_{p}$ and $n_{\mathsf {iter}}>1$.

\subsection{Pilot Design and Channel Estimation}
In this subsection, we first introduce the pilot pattern design. Then, our proposed pilot-aided channel estimation method is elaborated. Finally, the overhead for the different pilot patterns are compared.
\begin{figure*}[tbp]
\centering
\centerline{\includegraphics[width=1\textwidth]{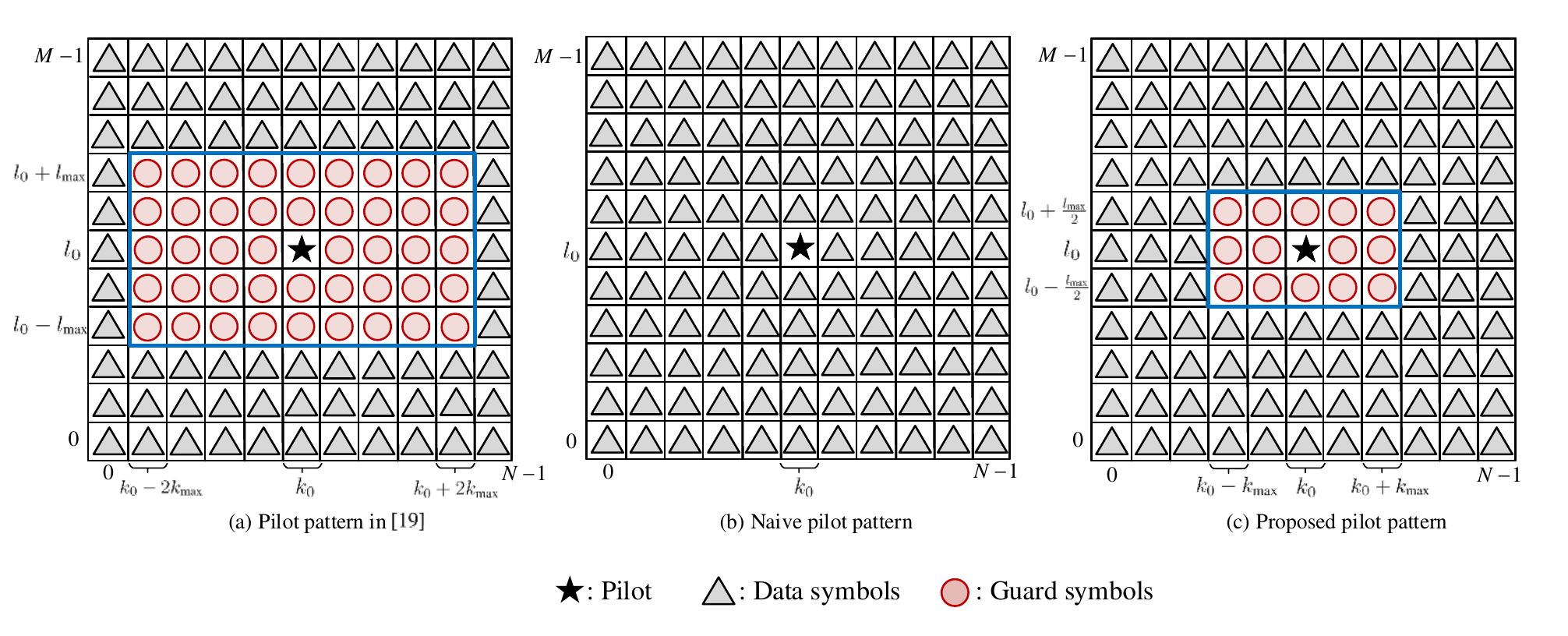}}
\caption{The comparison between the three pilot patterns with different overhead is as follows. The number of guard symbols in the pattern (c) is around $25$\% of that in \cite{CE4}.}
\label{fig}
\end{figure*}
\subsubsection{Pilot pattern design}
The pilot pattern in \cite{CE4} is depicted in Figure 5(a) and is formulated by
\begin{subequations}
\begin{numcases}{x[l,k]=}
d_{0},  \quad\quad\quad  {l=l_{0},\, k=k_{0}}     \\
0,      \quad\quad\quad\  {l \in [l_{0}-l_{\text{max}}, l_{0}+l_{\text{max}}]}\\
\quad   \quad\quad\quad  {k\in [k_{0}-2k_{\text{max}},k_{0}+2k_{\text{max}}]}\nonumber\\
d[l,k], \quad\ \, {\text{otherwise}}
\end{numcases}
\end{subequations}
where $d_{0}$ is the pilot symbol; $d[l,k]$ is the information-bearing symbols; the guard is $0$. $l_{\text{max}}=M\Delta f\tau_{\text{max}}$ and $k_{\text{max}}=NT\nu_{\text{max}}$ are the maximum delay and the maximum Doppler indices of the multipath channel, respectively. The position for the pilot is confined within the grid $l_{0}\in[l_{\text{max}},\ M-1-l_{\text{max}}],\ k_{0}\in[2k_{\text{max}},\ N-1-2k_{\text{max}}]$ as that in \cite{CE4} for ease of representation. From (\ref{eq:io2dim}), we see that the two-dimensional convolution spreads the data symbols and the pilot over the extent of the channel support in the delay-Doppler domain. The symbols go through all fading paths and are added up at the receiver. The essence of resolving the deconvolution in \cite{CE4} is keeping the pilot exclusively involving a two-dimensional convolution even after the channel spreading in the delay-Doppler domain. Thus, the guard symbols should be preserved in accordance with the scope of the channel support considering the data symbols spread which may contaminate the pilot.$\footnote{For instance, in terms of the Doppler dimension, the symbols at the position of $k_{0}+2k_{\text{max}}$ may be shifted to $k_{0}+k_{\text{max}}$ due to the Doppler effect in one path, while in another path, the pilot may be shifted from $k_{0}$ to $k_{0}+k_{\text{max}}$ at the same time. On this occasion, the pilot could be contaminated  by the data symbols in the literature, the range of the guard symbols along the Doppler dimension should be $[k_{0}-2k_{\text{max}},k_{0}+2k_{\text{max}}]$ to prevent this pollution. In contrast to the Doppler shift being either positive or negative, the delay is non-negative, and thus, the range of the guard in the delay dimension is $[l_{0}-l_{\text{max}},l_{0}+l_{\text{max}}]$. }$

The naive pilot pattern design without any guard in Figure 5(b) is expressed as
\begin{subequations}
\begin{numcases}{x[l,k]=}
d_{0},  \quad\quad\quad  {l=l_{0},\, k=k_{0}}     \\
d[l,k], \quad\ \, {\text{otherwise}}.
\end{numcases}
\end{subequations}
We have proved in (\ref{eq:dd1}) that with sufficiently many receive antennas, the signal from each path can be identified with trivial interference from other paths. For any signal from a single path, both the pilot and the data symbols experience the identical delay and Doppler shift, where the guard is unnecessary in this ideal case.
However, considering the finite receive antennas, the residual interference from the undesired directions may lead to the pilot and symbol blur. Thus, some guard symbols are still required to ensure accurate channel estimation.

We propose the pilot pattern in Figure 5(c) to strike a balance between the spectral efficiency and the accuracy of the channel estimation, and it can be formulated as
\begin{subequations}
\begin{numcases}{x[l,k]=}
d_{0},  \quad\quad\quad  {l=l_{0},\, k=k_{0}}     \\
0,      \quad\quad\quad\ {l \in \Big[l_{0}-1/2\ l_{\text{max}}, l_{0}+1/2\ l_{\text{max}}\Big]}\\
\quad   \quad\quad\quad  {k \in [k_{0}-k_{\text{max}}, k_{0}+k_{\text{max}}]}\nonumber\\
d[l,k], \quad\ \, {\text{otherwise}.}
\end{numcases}
\end{subequations}
In the simulation in Section IV, we show that the bit error rate (BER) using our pilot pattern is almost the same as that using the one in Figure 5(a).
\begin{figure*}[tbp]
\centering
\centerline{\includegraphics[width=1\textwidth]{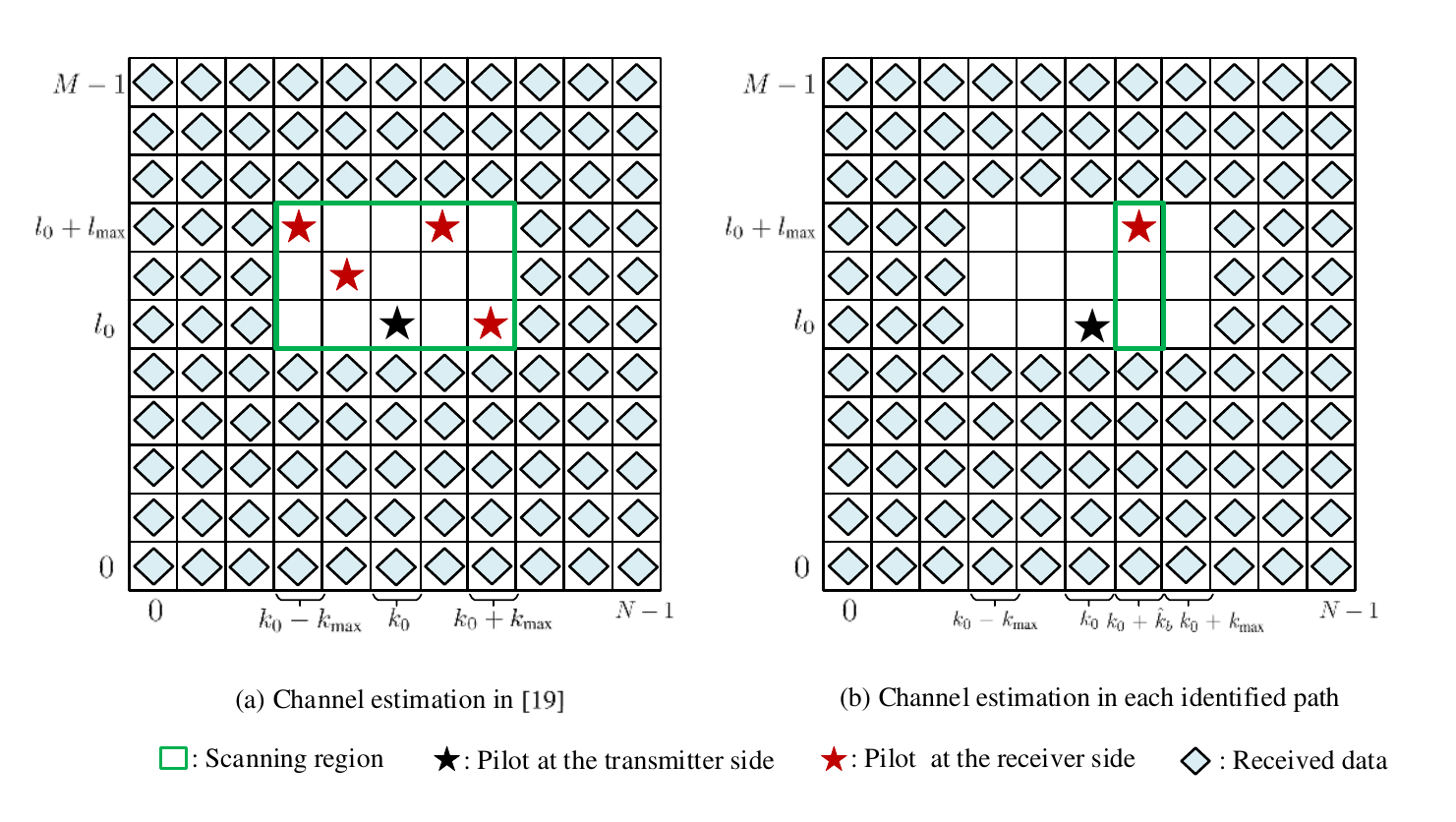}}
\caption{The comparison between the proposed channel estimation approach and the existing one in \cite{CE4}. In our approach in (b), the scanning region of each identified path is one dimension and only a single path requires to be detected, which decreases the probability of miss detection and false alarm in a noisy environment. In contrast, the scanning region (a) in \cite{CE4} is in the two dimensions and all paths of the channel are estimated in this region. }
\label{fig}
\end{figure*}

Furthermore, the overhead of our pilot pattern is much lower than the one in Figure 5(a). Specifically, the number of the symbol overhead is $(2l_{\text{max}}+1)(4k_{\text{max}}+1)$ in Figure 5(a). In contrast, the number is $(l_{\text{max}}+1)(2k_{\text{max}}+1)$ in the proposed pilot pattern and it is around $25$\% of that in Figure 5(a), which significantly decreases the overhead, especially in the high-speed scenarios.

\subsubsection{Low-overhead channel estimation}
The proposed channel estimation determines the Doppler shift, the delay, and the fading coefficient.
First, for an identified path, the Doppler shift is estimated by$\footnote{Here the index of the Doppler shift is rounded. The inaccuracy of the channel estimation due to this approximation diminishes as the large-scale antenna array provides a higher spatial resolution.}$
\begin{align}
\hat{k}_{b}&=\lfloor NTf_{\text{d}}\cos\varphi_{b}+0.5\rfloor,\qquad b \in\mathcal{I}_{B}.
\label{eq:Doppleresti}
\end{align}
Then, the delay can be determined by scanning the following region in the grid as
\begin{align}
\mathcal{D}_{b}=\{(l,k_{0}+
\hat{k}_{b} )|\ l\in[l_{0},l_{0}+l_{\text{max}}]\},\qquad b \in\mathcal{I}_{B}.
\end{align}
In Figure $6$, we provide a contrasting example of the search region. The search region of our approach leads to more accurate delay estimation and lower searching complexity. Thus, the delay index is estimated as
\begin{align}
\hat{l}_{b}=\arg\mathop{\max}_{l\in \mathcal{D}_{b}}\big|\bar{y}_{b}\big[l,k_{0}+
\hat{k}_{b}\big]\big|-l_{0},\qquad b \in\mathcal{I}_{B}.
\end{align}
Finally, with the estimates of the delay and the Doppler, the channel coefficient is estimated by
\begin{align}
\hat{\beta}_{b}={\frac{1}{d_{0}}}\bar{y}_{b}\big[l_{0}+\hat{l}_{b},k_{0}+\hat{k}_{b}\big]e^{\frac{j2\pi\hat{l}_{b}\hat{k}_{b}}{MN}},\qquad b \in\mathcal{I}_{B}.
\end{align}

\section{Simulation Results}
In this section, we evaluate the channel estimate performance and the error performance of the proposed receiver for OTFS with respect to (w.r.t.) different parameter settings. The channel estimation in \cite{CE4} and the equalization scheme in \cite{EQ4} are evaluated as the benchmark.
\begin{table}[tbp]
\setlength{\abovecaptionskip}{0pt}
\setlength{\belowcaptionskip}{0pt}\setlength{\abovecaptionskip}{0pt}
\setlength{\belowcaptionskip}{0pt}
\centering
\footnotesize
\newcommand{\tabincell}[2]{\begin{tabular}{@{}#1@{}}#2\end{tabular}}
\caption{Simulation Parameters}
\begin{tabular}{lll}
\toprule  
Parameters&Values\\
\midrule  
Number of OTFS symbols: $N$& $128$\\
Number of carriers: $M$&$512$\\
Carrier frequency: $f_{\rm{c}}$ &$4$ GHz\\
Subcarrier spacing: $\Delta f$&$15$ KHz\\
Distance of antenna elements: $\eta$&$0.45$ $\lambda$\\
Number of taps: $P$&$4$ and $6$\\
Delay power profiles                  &$P=4$:\,$[0, 370, 1090, 2510]$ (nsec);\\
&\quad\quad\quad$\,[0.0, -0.6, -7.0, -16.9]$ ($\mathrm{dB}$)\\
                               &$P=6$:\,$[0, 150, 370, 1090, 1730, 2510]$ (nsec);\\
&\quad\quad\quad$\,[0.0, -1.4, -3.6, -7.0, -12.0, -16.9]$ ($\mathrm{dB}$)\\
Modulation formats&$4$-QAM and $16$-QAM\\
Maximum delay index: $l_{\text{max}}$ &$20$\\
Number of receive antennas: $E$&$32$, $64$, $128$, $256$\\
Velocity: $v$ (km/hr)&$30$, $120$, $500$\\
\bottomrule 
\end{tabular}
\end{table}
\begin{figure*}[tbp]
\centering
\centerline{\includegraphics[width=5.5 in]{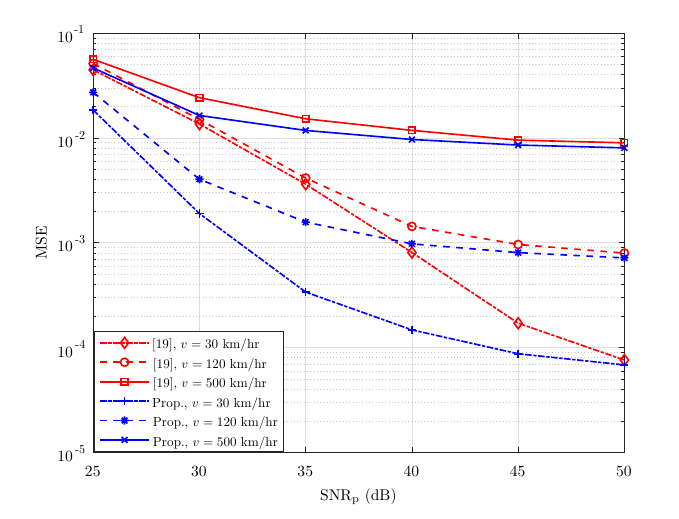}}
\caption{The MSE performance is evaluated for our channel estimation and the one in \cite{CE4} and $4$-QAM is adopted. The number of the receive antennas is $E=128$. Our channel estimation is more accurate with around $25$\% overhead of that in \cite{CE4}.}
\label{fig}
\end{figure*}

The simulation setup is listed in Table I. We first set $Q_{p}=1$, which aligns with that in \cite{EQ4}, \cite{EQ5}, and \cite{CE4}. Then, we provide the results for $Q_{p}>1$. The AoA of each path is independently uniformly distributed in $[0, 2\pi)$. In addition, the mean square error (MSE) of the channel estimation is evaluated. The SNR of the pilot $\text{SNR}_{\text{p}}$ is defined by $\mathsf{SNR}_{\text{p}}=\frac{|d_{0}|^{2}}{\sigma^{2}}$.


$\textbf{\emph{Observation 1:}}$
\noindent\emph{The proposed channel estimation outperforms the one in \cite{CE4} in different velocities, and it achieves $\mathit{5\ dB}$ gain at an MSE of $\mathit{10^{-3}}$ when the speed is over $\mathit{120}$ km/hr. (c.f. Figure 7)}

In Figure $7$, we evaluate the MSE performance in three different velocities and the SNR of the information-bearing data symbols is $20 \ \mathrm{dB}$. In addition, the channel estimation in \cite{CE4} serves as the benchmark, and $E=128$ is set for our proposed scheme. We can see that our proposed channel estimation achieves better MSE performance than the benchmark under the three moving speeds. Furthermore, the higher the pilot transmit power, the more accurate the channel estimation.
\begin{figure*}[tbp]
\centering
\centerline{\includegraphics[width=5.5 in]{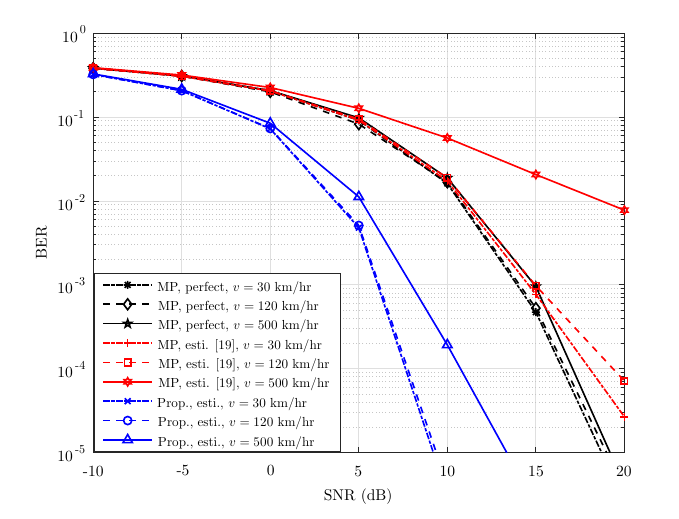}}
\caption{The BER is evaluated for our scheme and MP detection. The modulation format is $4$-QAM. The number of the receive antennas is $E=128$. The MP with the perfect channel knowledge is also evaluated. The proposed receiver achieves better error performance than the MP detection with either the estimated channel in \cite{CE4} or even the one with perfect channel knowledge.}
\label{fig}
\end{figure*}
This is because the received pilot is detected on the amplitude of the received signal, and high $\mathsf{SNR}_{\text{p}}$ increases the power disparity between the pilot and the data symbols. We can see that the improvement of the performance is not so obvious when $\mathsf{SNR}_{\text{p}}>40 \ \mathrm{dB}$. The following simulations are under $\mathsf{SNR}_{\text{p}}=40 \ \mathrm{dB}$.

$\textbf{\emph{Observation 2:}}$
\noindent\emph{The proposed receiver with the estimated channel achieves better BER performance in all three velocities than the MP detection either with the estimated channel in \cite{CE4} or even with the perfect channel. Moreover, the equalization complexity is reduced remarkably by the proposed receiver. (c.f. Figure 8 and Table II)}

\begin{table*}[tbp]
\setlength{\abovecaptionskip}{0pt}
\setlength{\belowcaptionskip}{0pt}\setlength{\abovecaptionskip}{0pt}
\setlength{\belowcaptionskip}{0pt}
\centering
\footnotesize
\newcommand{\tabincell}[2]{\begin{tabular}{@{}#1@{}}#2\end{tabular}}
\caption{Time Consumption of the Two Receivers}
\begin{tabular}{cccccc}
\toprule
\multicolumn{1}{c}{\multirow{2}*{Receivers}} &\multicolumn{5}{c}{Time consumption per frame (sec.)} \\
\cline{2-6}
\multicolumn{1}{c}{}&$0$ & $5$ & $10$ & $15$ & $20$\\
\hline
MP {[}12{]}&$413.167667$&$410.085087$&$409.070103$&$406.696567$&$23.242955$\\
Our proposed receiver&$0.960489$&$0.873761$&$0.903386$&$0.86901$&$0.9000855$\\
\bottomrule
\end{tabular}
\end{table*}

In Figure $8$, we compare the BER of our receiver with $E=128$ and the MP in the three velocities. In addition, the MP with perfect channel knowledge is also evaluated. We can see that our receiver achieves better error performance than the MP detection with either the estimated channel in \cite{CE4} or even the one with perfect channel knowledge. Furthermore, the performances of the MP with perfect channel knowledge under the three speeds are almost the same. Without the perfect channel information, the performances of our receiver at the speed of $30$ and $120$ km/hr are almost the same, and so is the MP with the estimated channel in \cite{CE4}. However, the performances for the two receivers become worse when $v=500$ km/hr which can be explained from the channel estimation performance in the velocity of $500$ km/hr.
\begin{figure*}[tbp]
\centering
\centerline{\includegraphics[width=5.5 in]{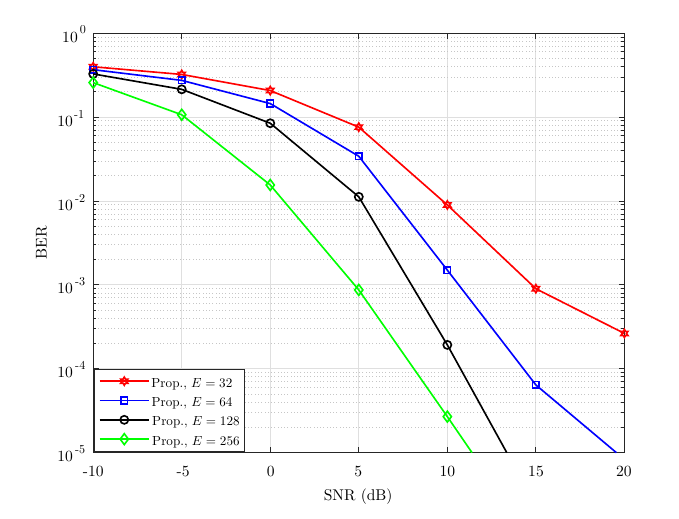}}
\caption{The BER performance of the proposed receiver is evaluated when $E=32$, $64$, $128$, and $256$. The modulation format is $4$-QAM, and the velocity is $v=500$ km/hr. It shows that the BER performance can be improved by increasing the number of the receive antennas. }
\label{fig}
\end{figure*}

The time consumption of the proposed equalization scheme and the MP is demonstrated in Table II. The time consumption is compared under $v= 500$ km/hr, and $E=128$ for our receiver. From Table II, we can see that the time consumption of our proposed scheme reduced to  $2.4$\textperthousand \ of MP in the low SNR region, and reduced to $4.3\%$ of MP under $\mathsf{SNR}=20\ \mathrm{dB}$. This is because, for our proposed scheme, the equalization proceeds as equalizing the received flat-faded signal without the iterative and the updating processes as MP. In addition, the loopy graph generates during the iterative procedure in the low SNR region for MP which increases the value of $n_{\mathsf {iter}}$. The time consumption of each scheme is evaluated in the same server with the identical setting. The trend of time computation is consistent with the complexity evaluation of the two receivers and confirms the lower complexity of the proposed scheme than the MP.

$\textbf{\emph{Observation 3:}}$
\noindent\emph{The BER performance of the proposed receiver can be improved by increasing the number of receive antennas. (c.f. Figure 9)}

\begin{table}[tbp]
\setlength{\abovecaptionskip}{0pt}
\setlength{\belowcaptionskip}{0pt}\setlength{\abovecaptionskip}{0pt}
\setlength{\belowcaptionskip}{0pt}
\centering
\footnotesize
\newcommand{\tabincell}[2]{\begin{tabular}{@{}#1@{}}#2\end{tabular}}
\caption{Channel Estimation Overhead for Different Patterns at the Speeds of $30/120/500$ km/hr.}
\begin{tabular}{ccccc}
\toprule
Pilot pattern &  \tabincell{c}{\# of pilot plus guards}&\tabincell{c}{\# of data symbols} &\tabincell{c}{Overhead percentage}\\
\hline
\multirow{1}*{\cite{CE4} in Figure 5(a)}&$205/697/2665$ &$65531/64839/62871$&$0.312\%/1.064\%/4.066\%$   \\
\multirow{1}*{Naive in Figure 5(b)}& $1/1/1$ &$65535/65535/65536$& $0.002\%/0.002\%/0.002\%$  \\
\multirow{1}*{Proposed in Figure 5(c)}& $63/189/693$ &$65473/65347/64843$& $0.096\%/0.288\%/1.057\%$ \\
\bottomrule
\end{tabular}
\end{table}

\begin{figure*}[tbp]
\centering
\centerline{\includegraphics[width=5.5 in]{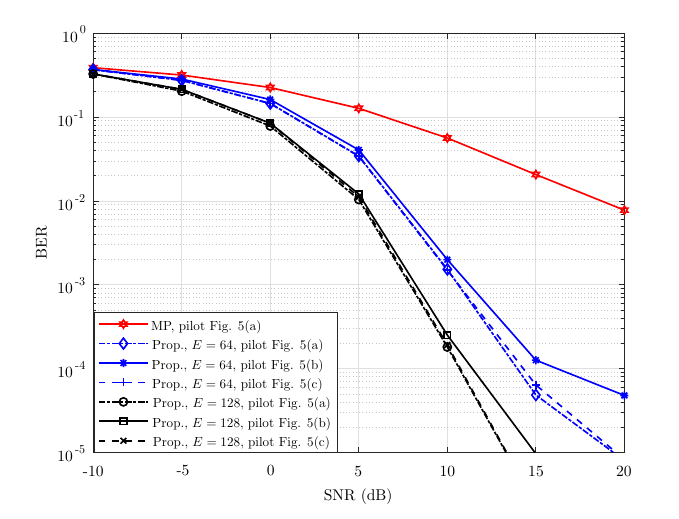}}
\caption{The BER is evaluated for the proposed receiver and the MP detection with the estimated channel. In addition, the performance of  the three pilot patterns is evaluated. The modulation format is $4$-QAM, and the velocity is $v=500$ km/hr. The proposed receiver outperforms the MP detection in \cite{CE4}. The proposed pilot pattern in Figure 5(c)  achieves sufficiently good BER performance as that of the pattern in Figure 5(a) of \cite{CE4}.}
\label{fig}
\end{figure*}

In Figure $9$, we demonstrate the BER of the proposed receiver when $E=32$, $64$, $128$, and $256$. We can see that with increasing the number of receive antennas, the error performance behaves better. This is because a higher spatial resolution is provided and the influence of the interference can be ignored with a sufficiently large $E$. Such better performance is at the cost of a large-scale antenna array. The large-scale antenna array can offer benefits such as high spatial resolution, high spectral efficiency, and so on. With the carrier frequency of a real communication system improved constantly, and the massive multiple-input and multiple-output technique deemed as one of the critical techniques of the fifth-generation systems, the cost of the large-scale antenna array is reduced, and the popularization of the large-scale antenna array is not out of reach \cite{MIMO1,MIMO2}.

\begin{figure*}[tbp]
\centering
\centerline{\includegraphics[width=5.5 in]{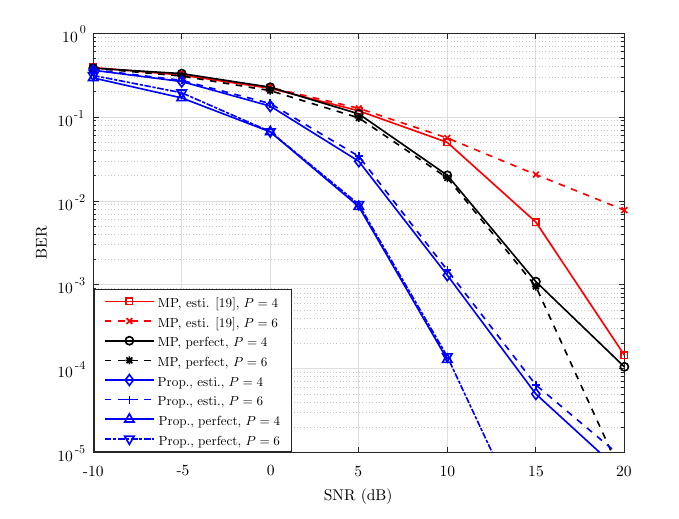}}
\caption{The BER is evaluated for the proposed receiver with $E=64$, and the MP detection when $P=4$, and $6$. The performance with the perfect channel knowledge is also evaluated. The modulation format is $4$-QAM, and the velocity is $v=500$ km/hr. The proposed receiver is robust to the variation of the numbers of the channel taps.}
\label{fig}
\end{figure*}
$\textbf{\emph{Observation 4:}}$
\noindent\emph{The proposed receiver achieves better BER performance with almost $\mathit{25\%}$ of the channel overhead in \cite{CE4}. (c.f. Table III and Figure 10)}

In Table III, we compare the overhead of the three pilot patterns including the patterns in Figure 5(a), Figure 5(b), and Figure 5(c) in three velocities. We can see that the channel estimation overhead increases with the terminal speed. The overhead of our proposed pattern in Figure 5(c) almost reduces to $25\%$ of the pattern in Figure 5(a). This indicates the low-overhead of the proposed pilot pattern.

\begin{figure*}[tbp]
\centering
\centerline{\includegraphics[width=5.5 in]{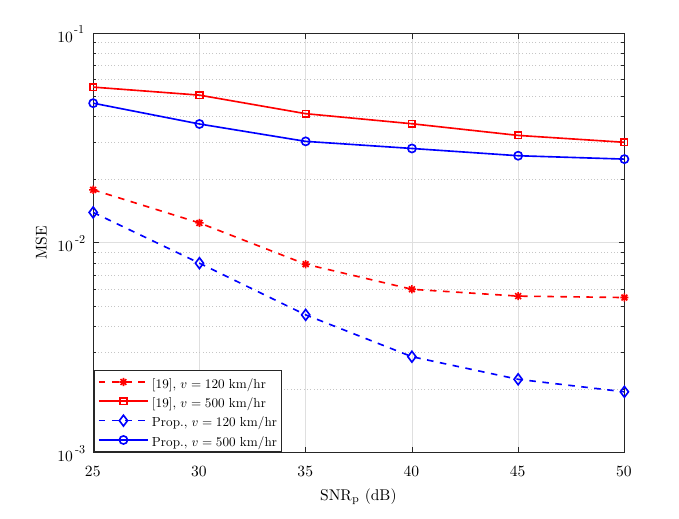}}
\caption{The MSE performance is evaluated for our channel estimation and the one in \cite{CE4} when a generalized multipath channel model is adopted. The number of the receive antennas is $E=128$, and the $4$-QAM is adopted. Our channel estimation is more accurate even under such a generalized channel model.}
\label{fig}
\end{figure*}

In Figure $10$, the BER is evaluated for our proposed receiver and the MP with the estimated channel at the speed of $500$ km/hr. In addition, the performances of our proposed scheme under the other two pilot patterns are evaluated. We can see that our proposed scheme outperforms the MP with the channel estimation in \cite{CE4}. Furthermore, the error floor emerges under the naive pilot pattern in Figure 5(b), since the interference originates from the non-orthogonality of the beamformers as can be seen from (\ref{eq:bf4}). Moreover, the pilot pattern in Figure 5(c) is sufficient to mitigate the interference, since the error performance under the pattern in Figure 5(c) is almost the same as that of the pattern in Figure 5(a). The proposed receiver achieves a better error performance with a lower pilot overhead, which is at the cost of a large-scale antenna array.

$\textbf{\emph{Miscellaneous observations:}}$
\emph{The proposed receiver achieves better performance under other parameter settings including (1) different numbers of the taps $P$; (2) the case of $\mathit{Q_{p}>1}$; (3) different modulation formats. (c.f. Figures 11-14)}

\begin{figure*}[tbp]
\centering
\centerline{\includegraphics[width=5.5 in]{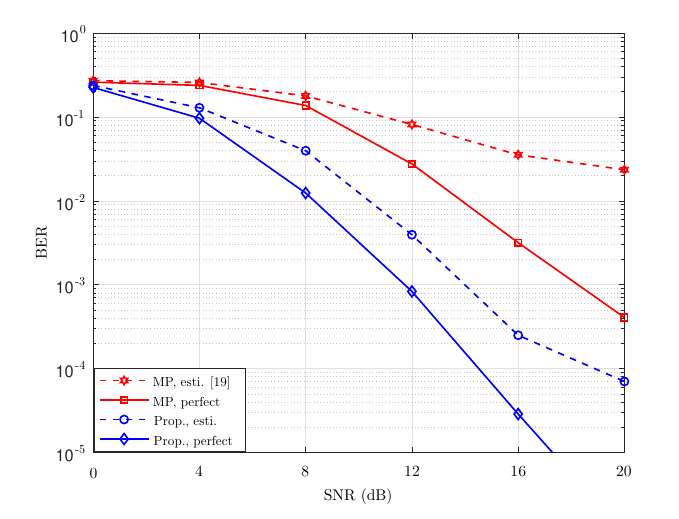}}
\caption{The BER is evaluated for the proposed receiver with $E=128$, and the MP when a generalized multipath channel model is adopted. The performance with the perfect channel knowledge is also evaluated. The modulation format is $4$-QAM, and the velocity is $v=500$ km/hr. The proposed receiver is applied to the generalized channel model and still outperforms the MP detection with either the estimated channel in \cite{CE4} or even the one with perfect channel knowledge.}
\label{fig}
\end{figure*}

In Figure $11$, we evaluate the BER for the proposed receiver and the MP when $P=4$ and $6$. The BER of the two receivers with and without the perfect channel knowledge is evaluated. The number of the receive antennas of our proposed receiver is $E=64$.
The modulation format is $4$-QAM, and the velocity is $v=500$ km/hr. We can see that our proposed receiver is robust to the variation of the taps number under both conditions.
This is because our proposed receiver does not rely heavily on the channel parameters and a high spatial resolution guarantees that all possible paths of the channel can be identified. The error performance becomes worse with increasing the number of taps for the channel estimation in \cite{CE4}.
For MP with the perfect channel knowledge, $P-1$ indicates the number of the considered interference terms. The high value of $P$, the more accurate of modeling the interference, and then the BER performance would be better \cite{CL}.

\begin{figure*}[tbp]
\centering
\centerline{\includegraphics[width=5.5 in]{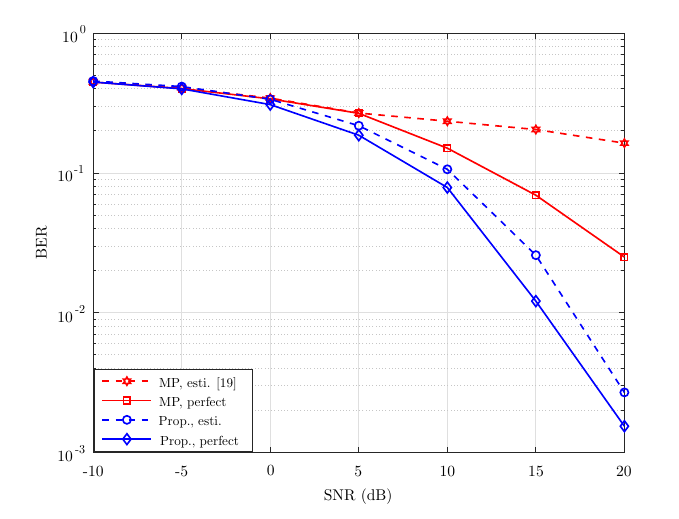}}
\caption{The BER is evaluated for the proposed receiver and the MP when the $16$-QAM modulation format is adopted. The performance with the perfect channel knowledge is also evaluated. The velocity is $v=500$ km/hr. Our scheme outperforms the MP detection with either the estimated channel in \cite{CE4} or even the one with perfect channel knowledge.}
\label{fig}
\end{figure*}

In Figure $12$, we evaluate the MSE performance in two velocities under the generalized channel model where more than one path corresponds to one tap, i.e., $Q_{p}>1$. The number of the taps is $P=6$, the total number of the paths is $\sum_{p}Q_{p}=16$. There are $2$, $3$, $4$, $3$, $2$, and $2$ paths for each tap correspondingly. The SNR of the information-bearing data symbols is $20 \ \mathrm{dB}$. In addition, the channel estimation in \cite{CE4} serves as the benchmark, and $E=128$ is set for our proposed scheme. We can see that our proposed channel estimation still achieves better MSE performance than the benchmark under the two moving speeds. This suggests that the proposed channel estimation scheme can be applied to the generalized channel model.

In Figure $13$, the BER is evaluated for our receiver and the MP under the above mentioned generalized channel model. The speed is $v=500$ km/hr, and the number of the receive antennas is $E=128$ for our proposed receiver.
We can see that under the generalized channel model the performance of the proposed receiver is better than the MP with the channel estimation in \cite{CE4} or even performs better than the MP with perfect channel knowledge. This demonstrates that our proposed receiver
can be extended to a generalized multipath channel model and still obtains better error performance.

In Figure $14$, we explore the performance of our proposed receiver and the MP under the $16$-QAM modulation format. The number of the receive antennas is $E = 128$ for our proposed receiver. In addition, the proposed receiver and the MP are both evaluated with and without perfect channel knowledge. The error performance of our proposed receiver is still better than the MP under the $16$-QAM modulation format.

\section{Conclusion}
In this paper, we proposed an OTFS receive approach regarding its practicality by using a large-antenna array. The marriage of the two techniques has been proved to be prosperous and fruitful, which is validated by our proofs and the simulation results. Compared to the schemes in the literature, the main advantages are summarized in three folds: 1) Reliability: the channel estimation is more accurate, and the BER performance has been improved in different speeds of the mobile users; 2) Efficiency: we can reduce the amount of the overhead to one-quarter of the existing one, and still guarantee the identical performance of the channel estimation and symbol detection; 3) Complexity: the involved two-dimensional convolution of the multipath channel fading and the information-bearing symbols in the OTFS is decoupled to multiple flat-faded transmissions, and the MRC is competent for the equalization instead of any approach to deconvoluting the received signal in the delay-Doppler domain. The ramification of this paper is that it could push the OTFS technique into practice by simply combining a large-scale antenna array, which becomes quite accessible currently in large carrier frequency. Future work may focus on the imperfect issues in our setup.
\renewcommand\thesection{\arabic{section}}
\begin{appendices}
\section{Asymptotic Analysis of Equation (\ref{eq:bf4})}
The asymptotic analysis of the interference term in (\ref{eq:bf4}) w.r.t. the number of the receive antennas is discussed here.
We first define the normalized array gain of a direction $G(\theta_{p',q'},\varphi_{b})$ by the ratio of the undesired array gain and the desired one, which is expressed as
\begin{align}
G(\theta_{p',q'},\varphi_{b})&\triangleq \frac{\big|\sum_{i=0}^{E-1}e^{j\phi_{i}\cos\theta_{p',q'}}w^{\ast}_{i}(\varphi_{b})\big|}{\big|\sum_{i=0}^{E-1}e^{j\phi_{i}\cos\varphi_{b}}w^{\ast}_{i}(\varphi_{b})\big|},\qquad \theta_{p',q'}, \varphi_{b}\in [0, 2\pi) ,\ b\in\mathcal{I}_{B}.
\label{eq:bf6}
\end{align}
The asymptotic analysis of the normalized array gain is given in Lemma $2$.
\begin{lemma}
For a large-scale uniform linear antenna array, the array gain from the undesired direction is zero in the limit of $E\rightarrow \infty$ as
\begin{align}
\lim_{E\rightarrow\infty}G(\theta_{p',q'},\varphi_{b})=0,\qquad \cos \theta_{p'q'} \neq \cos \varphi_{b},\ b\in\mathcal{I}_{B}.
\end{align}
\end{lemma}
\quad\quad $\emph{Proof:}$
The antenna gain from other undesired directions is further extended as
\begin{align}
G(\theta_{p',q'},\varphi_{b})&=\frac{\big|\sum_{i=0}^{E-1}e^{j\phi_{i}\cos\theta_{p',q'}}e^{-j\phi_{i}\cos\varphi_{b}}\big|}{\big|\sum_{i=0}^{E-1}e^{j\phi_{i}\cos\varphi_{b}}e^{-j\phi_{i}\cos\varphi_{b}}\big|}
\label{eq:bf7}\\
&=\frac{1}{E}\bigg|\sum_{i=0}^{E-1}e^{j\phi_{i}\left(\cos\theta_{p',q'}-\cos\varphi_{b}\right)}\bigg|
\label{eq:bf8}\\
&=\frac{1}{E}\bigg|\sum_{i=0}^{E-1}e^{j\frac{1}{\lambda}2\pi i\eta\left(\cos\theta_{p',q'}-\cos\varphi_{b}\right)}\bigg|
\label{eq:bf9}\\
&=\frac{1}{E}\Bigg|\frac{1-e^{j\frac{1}{\lambda}2\pi E\eta\left(\cos\theta_{p',q'}-\cos\varphi_{b}\right)}}{1-e^{j\frac{1}{\lambda}2\pi\eta\left(\cos\theta_{p',q'}-\cos\varphi_{b}\right)}}\Bigg|
\label{eq:bf10}\\
&=\frac{1}{E}\frac{\big|\sin\left(\frac{1}{\lambda}\pi E\eta\left(\cos\theta_{p',q'}-\cos\varphi_{b}\right)\right)\big|}{\big|\sin\left(\frac{1}{\lambda}\pi\eta\left(\cos\theta_{p',q'}-\cos\varphi_{b}\right)\right)\big|} ,\qquad b\in\mathcal{I}_{B}
\label{eq:bf11}
\end{align}
where $\cos \theta_{p'q'} \neq \cos\varphi_{b}$; (\ref{eq:bf7}) is obtained by plugging (\ref{eq:steer}) into (\ref{eq:bf6}); (\ref{eq:bf8}) is derived due to the denominator of (\ref{eq:bf7}) being E; (\ref{eq:bf9}) is obtained by plugging (\ref{eq:h3}) into (\ref{eq:bf8}); (\ref{eq:bf10}) is obtained by summing up the geometric sequence; (\ref{eq:bf11}) is obtained by applying the Euler's formula and some manipulation in the trigonometric function. Due to $\cos \theta_{p'q'} \neq \cos \varphi_{b}$ and then the denominator is nonzero, we can see that $G(\theta_{p',q'},\varphi_{b})$ approaches to zero for a sufficiently large $E$. Therefore, the proof of Lemma $2$ is completed.$\hfill\square$
\section{Proof of Lemma $1$}
The received signal in the frequency-time domain can be derived as
\begin{align}
\tilde{y}_{b}[m,n]&=\int_{t}g^{\ast}_{\text{r}}(t-nT)r_{b}(t)e^{-j2\pi m\Delta f(t-nT)}{\rm d}t
\label{eq:io2}\\
&=\int_{t}g^{\ast}_{\text{r}}(t-nT)\int_{-\nu_{\text{max}}}^{\nu_{\text{max}}}\int_{0}^{\tau_{\text{max}}}h_{b}(\tau,\nu)s(t-\tau)e^{j2\pi\nu(t-\tau)}{\rm d}\tau{\rm d}\nu\, e^{-j2\pi m\Delta f(t-nT)}{\rm d}t
\label{eq:io3}\\
&=\int_{t}g^{\ast}_{\text{r}}(t-nT)\int_{-\nu_{\text{max}}}^{\nu_{\text{max}}}\int_{0}^{\tau_{\text{max}}}\beta_{b}\delta(\tau-\tau_{b})\delta(\nu-\nu_{b}) e^{j2\pi\nu(t-\tau)}e^{-j2\pi m\Delta f(t-nT)}\nonumber\\
&\qquad \times \sum^{M-1}_{m'=0}\sum^{N-1}_{n'=0}\tilde{s}[m',n']g_{\text{t}}(t-\tau-n'T)
e^{j2\pi m'\Delta f(t-\tau-n'T)} {\rm d}\tau {\rm d}\nu{\rm d}t
\label{eq:io4}\\
&=\beta_{b}\int_{t}g^{\ast}_{\text{r}}(t-nT)e^{j2\pi\nu_{b}(t-\tau_{b})} e^{-j2\pi m\Delta f(t-nT)}
\nonumber\\
&\qquad \times\sum^{M-1}_{m'=0}\sum^{N-1}_{n'=0}\tilde{s}[m',n']g_{\text{t}}(t-\tau_{b}-n'T)e^{j2\pi m'\Delta f(t-\tau_{b}-n'T)} {\rm d}t
\label{eq:io5}\\
&=\beta_{b}\sum^{M-1}_{m'=0}\sum^{N-1}_{n'=0}\tilde{s}[m',n']\int_{t}g^{\ast}_{\text{r}}(t-nT)g_{\text{t}}(t-\tau_{b}-n'T)
\nonumber\\
&\qquad \times e^{j2\pi\nu_{b}(t-\tau_{b})} e^{-j2\pi m\Delta f(t-nT)}e^{j2\pi m'\Delta f(t-\tau_{b}-n'T)} {\rm d}t
\label{eq:io6}\\
&=\beta_{b}\sum^{M-1}_{m'=0}\sum^{N-1}_{n'=0}\tilde{s}[m',n']\int_{u}g^{\ast}_{\text{r}}\left(u-(n-n')T+\tau_{b}\right)g_{\text{t}}(u)\nonumber\\
&\qquad \times e^{j2\pi(u+n'T)\nu_{b}}e^{-j2\pi m\Delta f\left(u-(n-n')T+\tau_{b}\right)} e^{j2\pi m'\Delta fu}{\rm d}u
\label{eq:io7}\\
&=\beta_{b}\sum^{M-1}_{m'=0}\sum^{N-1}_{n'=0}\tilde{s}[m',n']\int_{u}g^{\ast}_{\text{r}}\left(u-(n-n')T+\tau_{b}\right)g_{\text{t}}(u) \nonumber\\
&\qquad \times e^{-j2\pi\left((m-m')\Delta f-\nu_{b} \right)\left(u-(n-n')T+\tau_{b}\right)}  e^{j2\pi(\nu_{b}+m'\Delta f)\left((n-n')T-\tau_{b}\right)}e^{j2\pi n'T\nu_{b}}{\rm d}u
\label{eq:io11}\\
&=\beta_{b}\sum^{M-1}_{m'=0}\sum^{N-1}_{n'=0}\tilde{s}[m',n']A_{g_{\text{r}},g_{\text{t}}}\big((m-m')\Delta f-\nu_{b},(n-n')T-\tau_{b}\big)\nonumber\\
&\qquad\times e^{j2\pi(\nu_{b}+m'\Delta f)\left((n-n')T-\tau_{b}\right)} e^{j2\pi n'T\nu_{b}}
\label{eq:io12}\\
&=\beta_{b}e^{j2\pi nT\nu_{b}}e^{-j2\pi(\nu_{b}+m\Delta f)\tau_{b}}\tilde{s}[m,n],\qquad b\in\mathcal{I}_{B}, \ m\in\mathcal{I}_{M}, \ n\in\mathcal{I}_{N}
\label{eq:io19}
\end{align}
where (\ref{eq:io2}) is obtained by the Wigner transform in (\ref{eq:modu5}); (\ref{eq:io3}) follows from (\ref{eq:modu3}); (\ref{eq:io4}) is obtained by plugging (\ref{eq:dd1}) and (\ref{eq:modu2}) into (\ref{eq:io3}); (\ref{eq:io5}) is derived due to the property of the Dirac delta function; (\ref{eq:io6}) is obtained by some manipulation of the order in (\ref{eq:io5}); (\ref{eq:io7}) is obtained by adopting the change of variable $u=t'-\tau_{b}-n'T$; (\ref{eq:io11}) is obtained by some algebraic calculations in (\ref{eq:io7}); (\ref{eq:io12}) is obtained by plugging (\ref{eq:cross}) into (\ref{eq:io11}); (\ref{eq:io19}) is derived due to the bi-orthogonality condition in (\ref{eq:ch3}).
Here, we complete the proof for the Lemma $1$.
\section{Proof of Theorem $1$}
The received symbol in the delay-Doppler domain can be expressed as
\begin{align}
\bar{y}_{b}[l,k]&=\frac{1}{\sqrt{MN}}\sum^{M-1}_{m=0}\sum^{N-1}_{n=0}\tilde{y}_{b}[m,n]e^{-j2\pi(\frac{nk}{N}-\frac{ml}{M})}
\label{eq:add2}\\
&=\frac{1}{\sqrt{MN}}\sum^{M-1}_{m=0}\sum^{N-1}_{n=0}\beta_{b}e^{j2\pi nT\nu_{b}}e^{-j2\pi(\nu_{b}+m\Delta f)\tau_{b}}\tilde{s}[m,n]e^{-j2\pi(\frac{nk}{N}-\frac{ml}{M})}
\label{eq:add3}\\
&=\frac{1}{MN}\sum^{M-1}_{m=0}\sum^{N-1}_{n=0}\beta_{b}e^{j2\pi nT\nu_{b}}e^{-j2\pi(\nu_{b}+m\Delta f)\tau_{b}}\sum_{l'=0}^{M-1}\sum_{k'=0}^{N-1}x[l',k']
 e^{j2\pi (\frac{nk'}{N}-\frac{ml'}{M})}e^{-j2\pi(\frac{nk}{N}-\frac{ml}{M})}
\label{eq:add4}\\
&=\frac{\beta_{b}e^{-j2\pi\tau_{b}\nu_{b}}}{MN}\sum_{l'=0}^{M-1}\sum_{k'=0}^{N-1}x[l',k']\sum_{m=0}^{M-1}e^{-j2\pi m\Delta f\tau_{b}}e^{j 2\pi m\frac{l-l'}{M}}\sum_{n=0}^{N-1}e^{j2\pi nT\nu_{b}}e^{-j2\pi n\frac{k-k'}{N}}
\label{eq:io24}\\
&=\frac{\beta_{b}e^{-j2\pi\tau_{b}\nu_{b}}}{MN}\sum_{l'=0}^{M-1}\sum_{k'=0}^{N-1}x[l',k']\sum_{m=0}^{M-1}e^{-j2\pi \frac{m}{M}\left(l_{b}-(l-l')\right)}\sum_{n=0}^{N-1}e^{j2\pi \frac{n}{N}\left(k_{b}-(k-k')\right)},
\label{eq:add5}\nonumber\\
&\qquad\qquad\qquad\qquad\qquad\qquad\qquad\qquad\qquad\qquad\qquad b\in\mathcal{I}_{B}, \ l\in\mathcal{I}_{M}, \ k\in\mathcal{I}_{N}
\end{align}
where (\ref{eq:add2}) is obtained by utilizing the SFFT in (\ref{eq:modu6}); (\ref{eq:add3}) follows from (\ref{eq:io19}); (\ref{eq:add4}) follows from (\ref{eq:modu1});  (\ref{eq:io24}) is obtained by some manipulation of the order in (\ref{eq:add4}); (\ref{eq:add5}) follows from the fact that $l_{b}=M\Delta f\tau_{b}$ and $k_{b}=NT\nu_{b}$.
For the geometric sequence $\sum_{m=0}^{M-1}e^{-j2\pi \frac{m}{M}\left(l_{b}-(l-l')\right)}$ in (\ref{eq:add5}), we can find that $\sum_{m=0}^{M-1}e^{-j2\pi \frac{m}{M}\left(l_{b}-(l-l')\right)}=M$ when the common ratio $e^{-j2\pi \frac{1}{M}\left(l_{b}-(l-l')\right)}=1$, then $[l_{b}-(l-l')]_{M}=0$; otherwise, the sum for the sequence is $\frac{1-e^{-j2\pi\left(l_{b}-(l-l')\right)}}{1-e^{-j2\pi \frac{1}{M}\left(l_{b}-(l-l')\right)}}=0$, due to the numerator is $0$, and the denominator is nonzero. Since $l_{b},\ l,\ l'\in\mathcal{I}_{M}$, only the single term $l'=[l-l_{b}]_{M}$ is considered for the sum for $l'$. Similarly, $\sum_{n=0}^{N-1}e^{j2\pi \frac{n}{N}\left(k_{b}-(k-k')\right)}=N$, if and only if $[k_{b}-(k-k')]_{N}=0$; otherwise, the sum for the sequence is $\frac{1-e^{j2\pi\left(k_{b}-(k-k')\right)}}{1-e^{j 2\pi\frac{1}{N}\left(k_{b}-(k-k')\right)}}=0$. Only the single term $k'=[k-k_{b}]_{N}$ is considered for the sum for $k'$ with the reason that $k_{b},\ k,\ k'\in\mathcal{I}_{N}$. Combining the result in (\ref{eq:add5}), the input-output relation can be further expressed as
\begin{align}
\bar{y}_{b}[l,k]&=\beta_{b}e^{-j2\pi\tau_{b}\nu_{b}}x[[l-l_{b}]_{M},[k-k_{b}]_{N}],\qquad b\in\mathcal{I}_{B},\ l\in\mathcal{I}_{M}, \ k\in\mathcal{I}_{N}.
\end{align}
Therefore, the proof of Theorem $1$ is concluded.
\end{appendices}

\bibliographystyle{IEEEtran}
\bibliography{REF}

\end{document}